# Inertia and scaling in deterministic lateral displacement

**Timothy J. Bowman[1] German Drazer[2,3] and Joelle Frechette[1,3]**

1. Department of Chemical and Biomolecular Engineering, Johns Hopkins University, 3400 N. Charles St., Baltimore MD 21210, USA
2. Department of Mechanical and Aerospace Engineering, Rutgers University, 98 Brett Rd, Piscataway, NJ 08854-8058
3. Author to whom correspondence should be addressed: german.drazer@rutgers.edu and jfrechette@jhu.edu

The ability to separate and analyze chemical species with high resolution, sensitivity, and throughput is central to the development of microfluidics systems. Deterministic lateral displacement (DLD) is a continuous separation method based on the transport of species through an array of obstacles. In the case of force-driven DLD (*f*-DLD), size-based separation can be modelled effectively using a simple particle-obstacle collision model. We use a macroscopic model to study *f*-DLD and demonstrate, via a simple scaling, that the method is indeed predominantly a size-based phenomenon at low Reynolds numbers. More importantly, we demonstrate that inertia effects provide the additional capability to separate same size particles but of different densities and could enhance separation at high throughput conditions. We also show that a direct conversion of macroscopic results to microfluidic settings is possible with a simple scaling based on the size of the obstacles that results in a universal curve.

## I. INTRODUCTION

Deterministic lateral displacement (DLD) is a promising microfluidic separation method in which different constituents of a mixture travel in different directions through an array of obstacles.[1-10] It has, in fact, been successfully used in a number of applications, especially for the separation of biological samples.[11-18] Although in most cases DLD is described (and investigated) as a size-based separation technique, it could also separate particles based on shape and flexibility[19, 20] and may contribute to the growing field of chiral separation.[21, 22]

We have shown that a simple model[23, 24] that describes the motion of a suspended sphere around an individual obstacle in the DLD array (which we refer to as a *particle-obstacle collision)* predicts the separation capability of the system.[24-28] In this model, two types of particle–obstacle collisions are considered in the low Reynolds number limit (see Fig. 1). The first, purely hydrodynamic collisions, produces no net change in the particle motion (top trajectory in Fig. 1). In particular, the upstream or incoming offset in the trajectory of the particle with respect to the centre of the obstacle ($b_{in}$ in Fig. 1) is the same as the downstream or outgoing offset after the collision ($b_{out}$ in Fig. 1). The second, touching collisions, are irreversible and give rise to a net lateral displacement in the trajectory followed by the particles (bottom trajectory in Fig. 1). According to the model, the type of collision depends on the incoming offset ($b_{in}$) and the resulting minimum separation gap between surfaces as the particle moves past the obstacle. Irreversible collisions occur for incoming offsets that are below some critical value, $b_c$ (corresponding to the middle trajectory in Fig. 1). It is the cumulative effect of such irreversible collisions that leads to particles migrating away from the direction of the force that drives them through the DLD array. Differences in the critical offset ($b_c$) of the different species leads to their different migration angles, thus resulting in separation. It is therefore a key parameter in the design and optimization of DLD separation systems.

Recently, we have derived rigorous results that relate the incoming offset to the minimum separation distance between the obstacle and the particle surfaces reached during the *collision* in the low Reynolds number limit [23] and explored inertia effects in numerical simulations[29]. Little is known, however, about what determines $b_c$ for a given mixture of particles to be fractionated. This critical offset can depend on particle sizes and materials but also on the characteristics of the separation system, such as obstacle size and material, fluid properties and driving field. The critical offset  is essentially a lumped parameter that encompasses multiple and likely coupled effects such as surface roughness, electrostatic repulsion, van der Waals attraction, particle inertia, and multibody hydrodynamic interactions with neighbouring obstacles.

Here, we study the dependence of the critical offset on particle-obstacle size-ratio, for spherical particles of different sizes and materials, in the case of force-driven DLD (*f*–DLD). We show that in the absence of inertia, *f*-DLD is predominantly a size-based phenomenon (we do not consider different particle shapes, which could also provide a basis for separation in the absence of inertia). More important, we show that *f*-DLD can separate particles of the same size based on density differences in the presence of particle inertia, a desirable feature that could provide/enhance separation at high throughput conditions. Finally, we demonstrate for the first time that results obtained in macroscopic models provide valuable information on the microfluidic system. More specifically, the results obtained in microfluidic systems agree with macroscopic results when rescaled by the size of the obstacle.

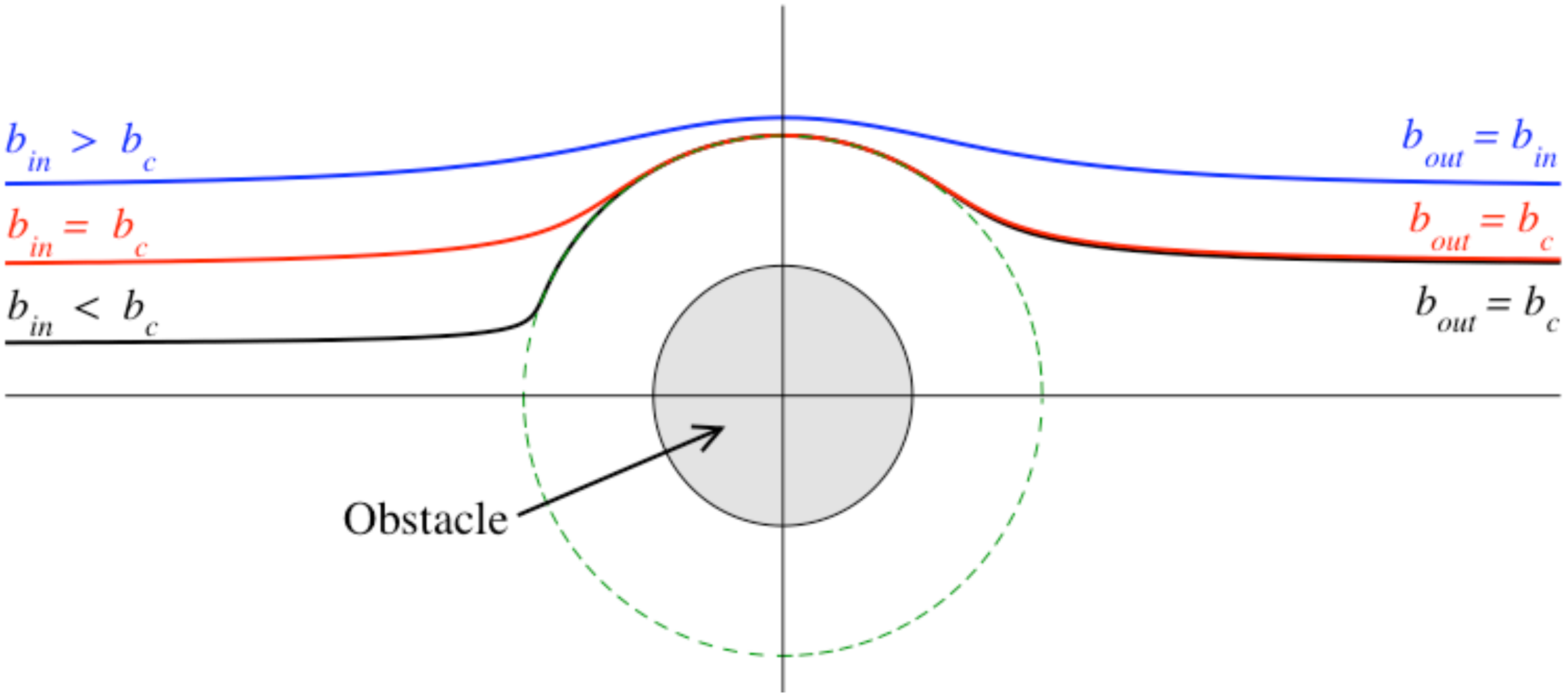

FIG 1. Illustration of the collision model showing the reversible and irreversible particle trajectories around a fixed obstacle (from Ref. [1]). Reversible, purely hydrodynamic collisions are those for which bin > bc and lead to bout = bin. Irreversible collisions are those for which $b_{in} < b_c$ and lead to $b_{out} = b_c$. The dotted line around the obstacle indicates the excluded volume.

## II. MATERIALS AND METHODS

### A. Experimental setup and materials

We use a macroscopic model system to study the critical offset to characterize the separation of particles in *f*-DLD. There are several benefits of using macroscopic models, such as faster experimentation, lack of particle deposition and fouling, high design fidelity, and the possibility to probe a truly deterministic regime. The macroscopic model employed here consists of cylindrical LEGO® posts of diameter $2R$ = 7.8 mm (*small obstacles,* shown in Fig. 2) or $2R$ = 15.8 mm (*large obstacles*). The obstacles are snapped onto a LEGO® board (measuring 305 mm × 305 mm) that is glued to a 0.5 inch thick acrylic backing to improve the rigidity of the board. Two stacked cylindrical LEGO® pegs constitute one obstacle of height 19.1 mm. We use two pegs to increase the distance between the sedimenting spheres and the board in our experiments, thus reducing any possible effect of its protrusions. We use two different configurations: (i) a square array of obstacles and (ii) individual columns of equally spaced obstacles (with column–column spacing of at least 110 mm, as shown in Fig. 2). The center–to–center distance between two obstacles, $l$, is varied between 7.8 mm and 24.2 mm. The gap between obstacles, $\Delta$, is simply $l - 2R$. We also define the gap-to-obstacle aspect ratio $\delta = \Delta/2R$. The board is immersed in a large optically transparent acrylic tank filled with either 55 cSt soybean oil (*low viscosity fluid*) or 405 cSt silicone oil (*high viscosity fluid*). We rotate the board to vary the forcing angle ($\theta$), which is the angle formed between a column of obstacles and the direction of the driving force (here gravity). Therefore, a forcing angle of $\theta = 0°$ corresponds to a vertical column of obstacles. The materials for the spherical particles are nylon ($\rho = 1.084$ g/cm$^3$), poly(methyl methacrylate) (PMMA) ($\rho = 1.188$ g/cm$^3$), Delrin® ($\rho = 1.410$ g/cm$^3$), borosilicate glass ($\rho = 2.230$ g/cm$^3$, aluminium 2017 ($\rho = 2.700$ g/cm$^3$), and 440C stainless steel ($\rho = 7.650$ g/cm$^3$), all from McMaster Carr. The particles range in size ($2a$) from 1.0 mm to 12.3 mm in diameter with nominal deviations in sphericity less than 0.01%.

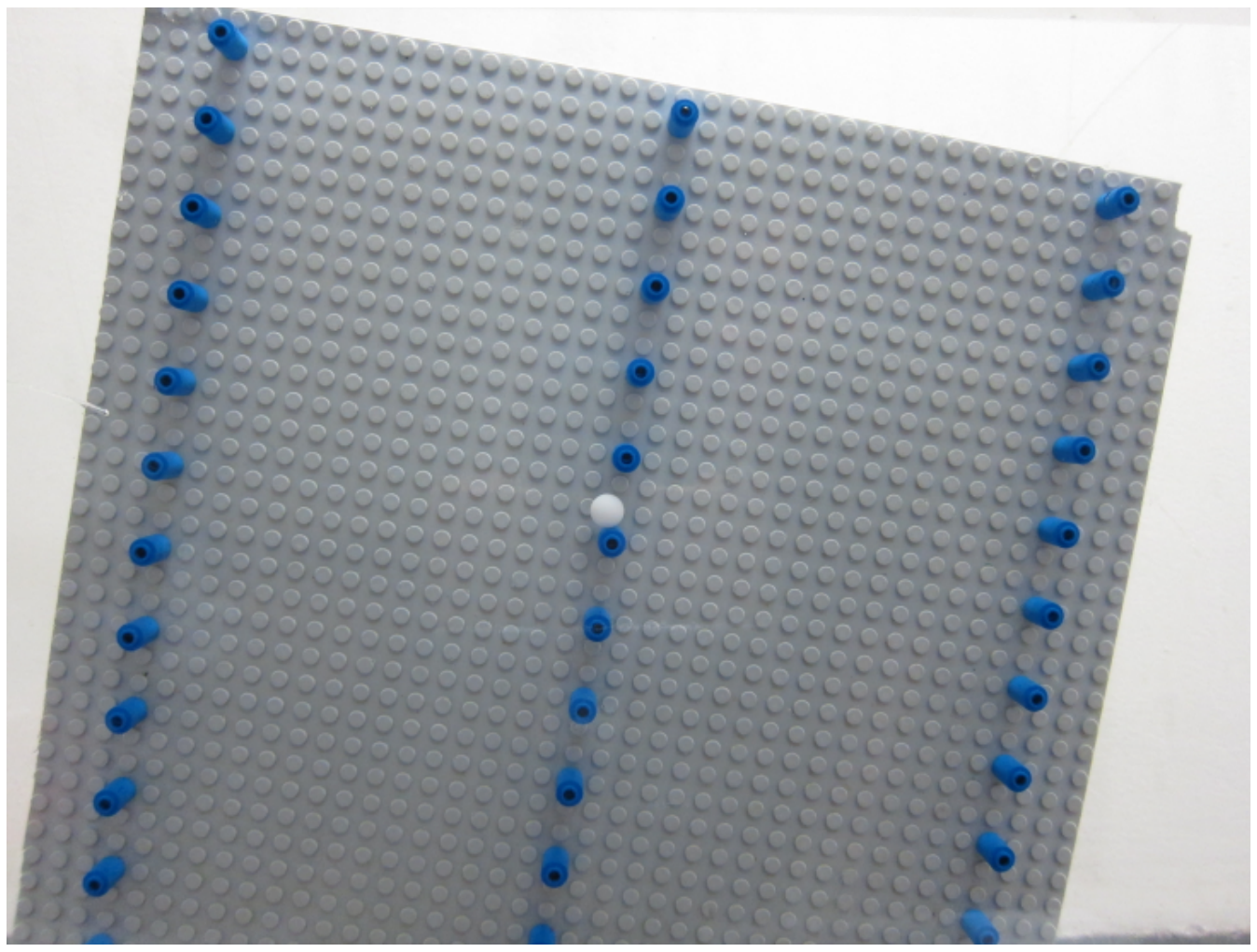

FIG.2. Picture of the experimental setup. In this case, the LEGO® obstacles of diameter $2R = 7.9$ mm have a spacing $l = 24.0$ mm within a column. A 9.5 mm diameter white Delrin® particle is shown falling down the central column oriented at a forcing angle of $\theta = 9.8°$.

## B. Collision model

We employ a simple particle-obstacle collision model to characterize the trajectory of a spherical particle past a fixed cylindrical obstacle.[23, 24] This model considers individual particle-obstacle interactions (dilute limit). Therefore, long-range hydrodynamic interactions caused by neighboring obstacles (or other suspended particles) are neglected, and the complete motion of an individual particle is treated as a series of independent particle-obstacle collisions. Note that due to screening effects[30-32] hydrodynamic interactions between the suspended particle and an obstacle (or another particle) can be neglected when the separation distances are significantly larger than the space between the two flat plates enclosing the board (≈20mm).[33] The model treats each particle-obstacle collision as either a purely hydrodynamic collision or a touching collision, depending on the incoming offset or, equivalently, the minimum particle-obstacle separation reached along the trajectory (see Fig. 1). In the absence of inertia, purely hydrodynamic collisions result in symmetric particle trajectories around the obstacle, such that the incoming ($b_{in}$) and outgoing ($b_{out}$) offsets are the same. The offset is defined as the perpendicular distance from a line parallel to the force that goes through both the obstacle and the particle center. Irreversible or touching interactions, on the other hand, are asymmetric and lead to a net lateral displacement. The asymmetric trajectories resulting from touching collisions can be described using a hard-core repulsive potential that impedes the suspending particles from reaching separations smaller than the range of the repulsive core, ε, but has no effect as the particles moves away from the obstacle. The offset for which the minimum separation reached along the trajectory is exactly ε is called the critical offset or $b_c$ (see Fig. 1). Touching collisions, then, occur whenever the incoming offset is less than $b_c$. Since all touching collisions result in the same minimum surface separation between the particle and obstacle, ε, all of the corresponding outgoing trajectories collapse into a single trajectory with an outgoing offset equal to $b_c$. Therefore, hydrodynamic collisions are those for which $b_{in} > b_c$ leading to $b_{out} = b_{in}$, whereas irreversible touching collisions are characterized by $b_{in} \leq b_c$ and $b_{out} = b_c$ (see Fig. 1). The magnitude of $b_c$, or similarly $(a + R - b_c)$, gives an indication of how much a particle comes back around the obstacle on the receding side of the trajectory due to hydrodynamic interactions.

### C. First critical angle measurements

We measure the first critical angle ($\theta_c$), which is the smallest angle at which a particle crosses a column of obstacles (see Fig. 3). To obtain $\theta_c$ we monitor the trajectory of the particles as they settle past a single column of obstacles. Working with a single column of obstacles (instead of a full array) simplifies the experiments[1, 25] and we have shown previously, although using liquid drops, that the first critical angle obtained from a full obstacle array is the same as for a single column.[25] We have also validated this result for solid particles as shown in the Supplementary Information[34] (see Fig. S1). It is important to note that our previous work has shown that the first critical angle is the most sensitive to particle size, and therefore the design of a *f*-DLD system for size-based separation would typically rely on the first critical angles of the different components.

In the experiments, starting from the columns oriented vertically, we increase the forcing angle in small increments (rotating the board in 0.3° steps) and follow three particles as they settle past the column for each angle. We then estimate the critical angle as the first angle at which all three particles cross the column (at different obstacles). The reported value is the average of the critical angle measured in the three different columns shown in Fig. 2.

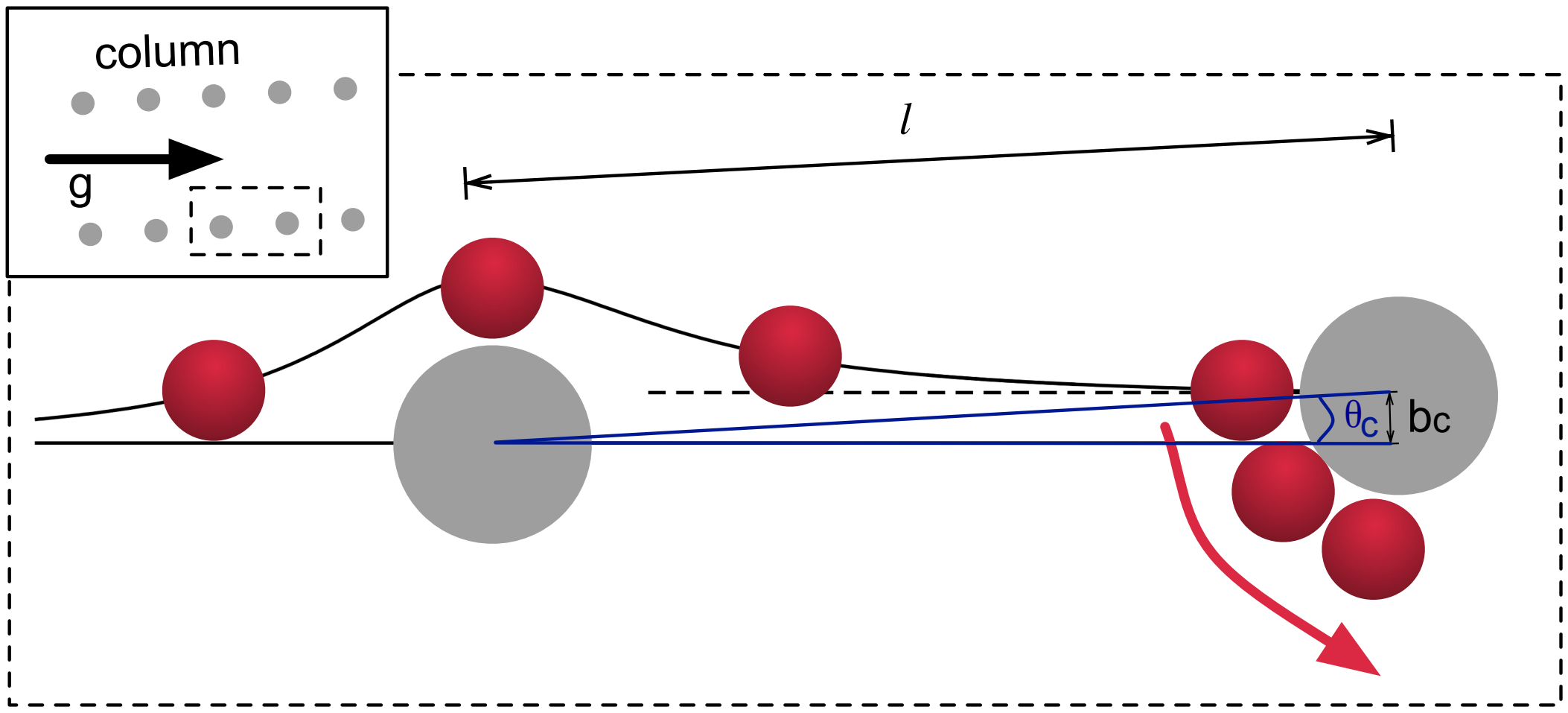

FIG. 3. Schematic of how the critical offset is evaluated from the first critical angle.

The collision model is then employed to calculate the critical offset ($b_c$) using the simple geometric relation $b_c = l\, sin(\theta_c)$. This relation is derived by comparing the lateral displacement induced by a touching collision with the lateral shift between successive obstacles in a column with respect to the direction of the driving force, as shown in Fig. 3. When the forcing angle is less than the critical angle, the particle will move along a column of obstacles. Only for $\theta > \theta_c$ the particle crosses the column as schematically shown in Fig. 3 (at the critical angle).

As discussed in the appendix, we also observed that increasing the separation between obstacles within a column led to a small decrease in the critical offset for large particles. This effect is due to the fact that the proximity between neighbouring obstacles prevents a particle from reaching its asymptotic offset. Therefore, the experimental $b_c$ value obtained from the measured critical angle is not the asymptotic bc, but instead corresponds to an effective critical offset that is a function of the lattice spacing.

## III. RESULTS AND DISCUSSION

### A. Material independence and inertia effects

In Fig. 4 we plot the first critical angle and associated $b_c$ as a function of particle diameter for six different materials. It is clear that, although there are some differences depending on the particle material, the dependence of the critical angle on particle size for particles with the lowest densities (Nylon, PMMA and Delrin) is similar. This indicates that the observed differences in particle trajectories for the higher density materials (aluminum, glass, and steel) could be due to inertia effects. In fact, for a given particle size increasing values of $b_c$ correspond to materials of increasing density. We define the particles Reynolds ($Re = \frac{\rho_f U d}{\mu}$) and Stokes ($St = \frac{1}{9}\left(\frac{\rho_s}{\rho_f}\right) Re$) numbers in terms of the average settling velocity (U) of the particles where $\rho_f$ is the fluid density, $d$ is the particle diameter, and μ is the viscosity of the fluid. The Stokes numbers range from 0.01 to 30 (see the numbers alongside the data points in Fig. 4). We note that we have observed some particles moving at relatively large Stokes numbers (St > 10) slightly bounce during a collision (at relatively small angles). However, we do not observe bouncing near the critical angles for the particles considered here.

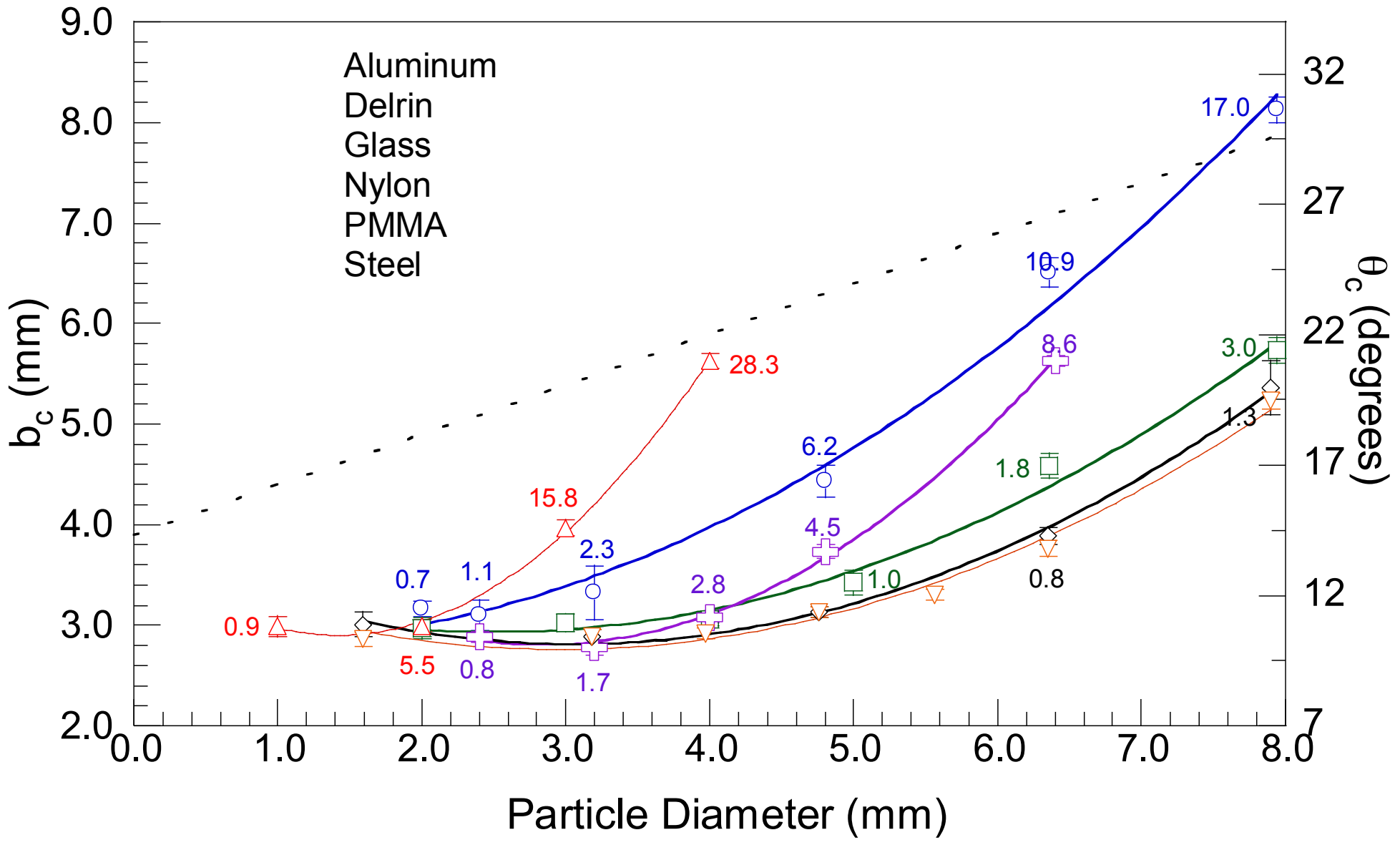

FIG. 4. Critical offset (left axis) and angle (right axis) as a function of particle size. (R = 3.9 mm; l = 16.0 mm). Lines are fitted to the data to guide the eye. The dashed line corresponds to bc = a+R. Stokes numbers greater than 0.5 are indicated.

To decouple materials properties from inertia effects we performed additional experiments using a high viscosity fluid. Shown in Figure 5 is the significant decrease in the critical offset for steel and aluminum particles when the Stokes number is decreased by using the more viscous fluid. It is also clear from Fig. 5 that, surprisingly, the critical offset becomes independent of the material at low Stokes numbers. This is a remarkable result, which suggests that inertia effects might be desirable in order to separate particles of different materials. Another important observation is that small particles show little dependence of the critical offset on size, which would make size-based separation difficult. Moreover, there is a slight increase in the critical offset for the smallest particles, which is consistent with previous simulations,[23, 29] and could complicate size-separation. However, we shall show in the next section that the range of particle sizes with a nearly constant critical offset depends on the size of the obstacles and can be avoided using obstacles that are smaller than the smallest particles to be fractionated.

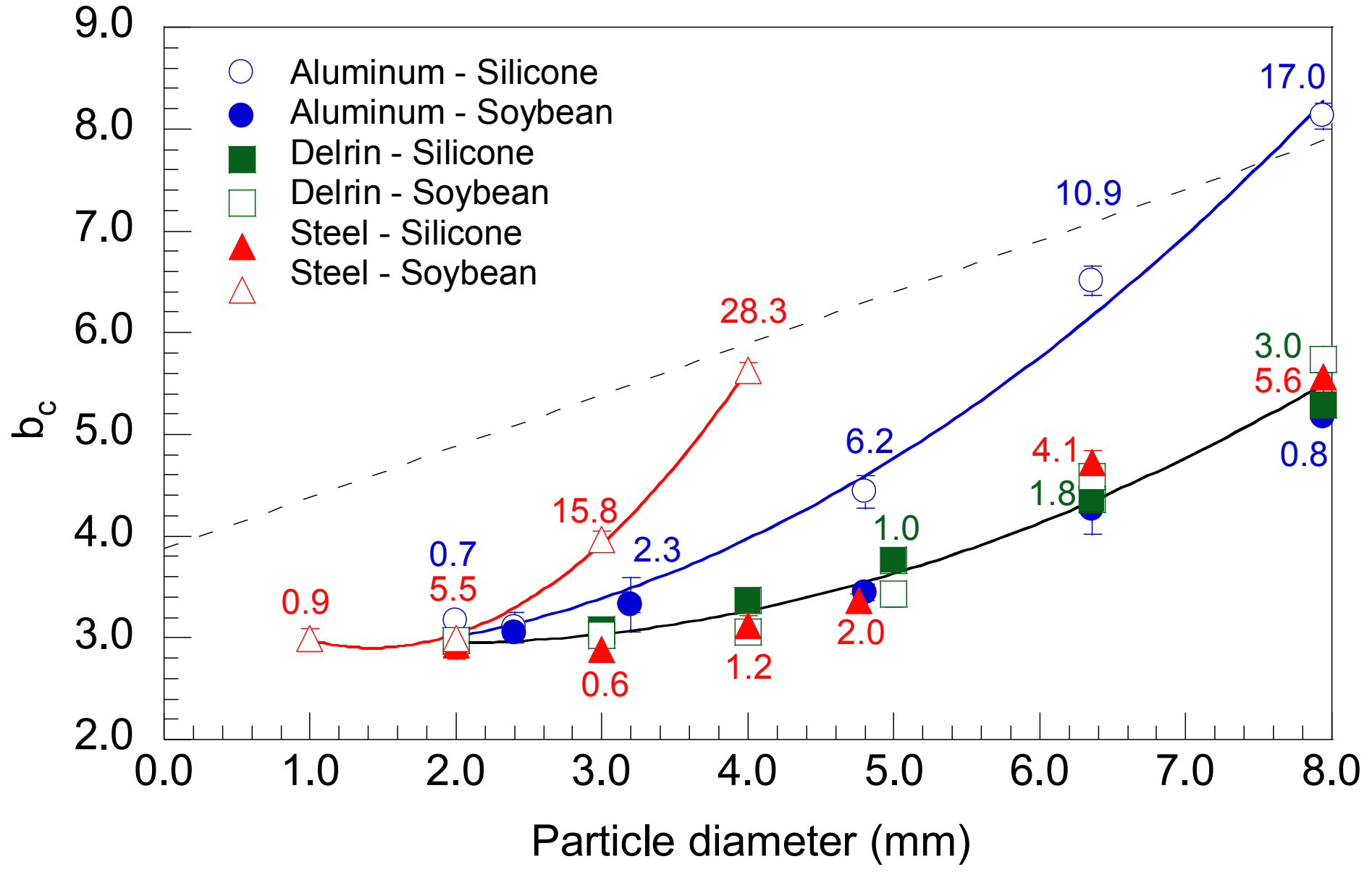

FIG. 5. Critical offset as a function of particle size (using two fluids of different viscosities). (R = 3.9 mm; $l$ = 16.0 mm; same configuration as in Fig. 4). Stokes numbers greater than 0.5 are indicated. Lines are fitted to the data to guide the eye. The dashed line represents $b_c = a + R$.

## B. Scaling behavior

For the previous results to be valid for the design of microfluidic DLD devices they have to apply when the size of the system is rescaled. Therefore, we investigate the scaling behaviour of the effective critical offset ($b_c$). In particular, we performed experiments in which we doubled the size of the obstacles and lattice spacing (we used obstacles with a diameter of 15.8 mm, which is approximately twice the size of the obstacles used in the previous experiments). First, in Fig. 6a we show the critical offset as a function of particle size for both systems. Then, in Fig. 6b we show that the results actually scale with the size of the system. That is, we obtain nearly identical behavior for the non-dimensional critical offset as a function of the non-dimensional particle size, using the size of the obstacles (R) as the characteristic length in each system. Again, this is a remarkable and unexpected result. From purely hydrodynamic considerations, we would expect that two systems that satisfy geometric similarity would have the same non-dimensional critical offset, $b_c/R$, but only provided that the non-dimensional roughness (or non-dimensional range of non-hydrodynamic forces) is also the same. This result also suggests that using relatively small obstacles ($a/R > 1$) would enhance separation resolution as the dependence of size is significantly larger in this region.

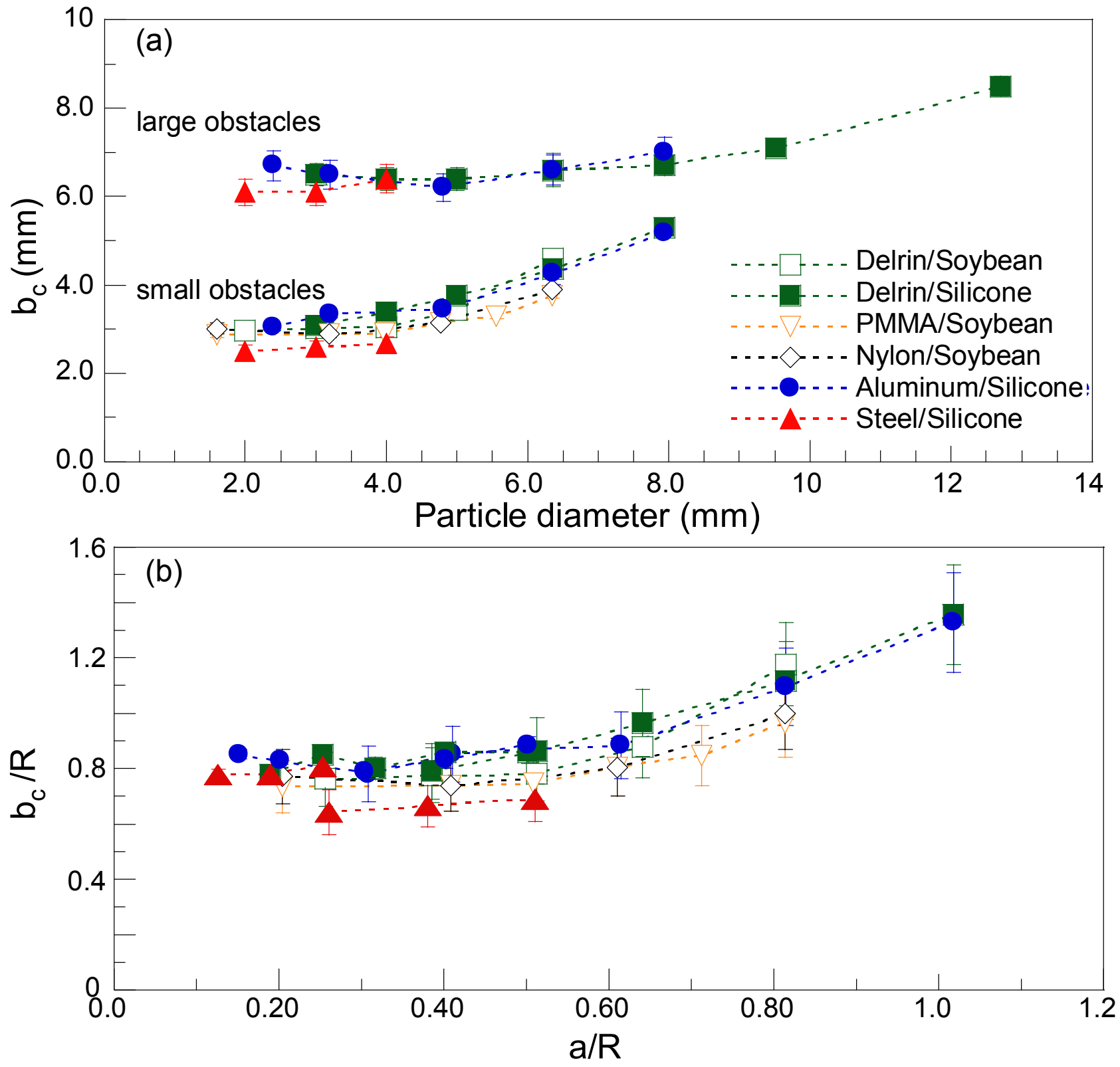

FIG. 6. (a) Critical offset, $b_c$, for two different obstacles and different particle materials. The small obstacle has a diameter of 7.9 mm and the large obstacle has a diameter of 15.8 mm. (b) Scaling of the critical offset and the particle radius with the obstacle radius (R) showing collapse of the data. In both plots we present data corresponding to small Stokes numbers (St < 5).

## C. Discussion and comparison with simulations and microfluidic data

In Fig.7, we compare the experimental results with our model predictions for different non-dimensional values of the range of the repulsive force (or effective roughness). The simulations are based on a spherical particles moving past a fixed sphere with a repulsive core of dimensionless range $\varepsilon$ (see Ref. [24] for details). When comparing the simulations to the experiments we use experimental data obtained using Delrin® particles and small obstacles, with large obstacle-obstacle separation ($\delta = 3$) to reduce multi-body hydrodynamic interactions (see Appendix). We see in Fig 7 that our experiments fall between non–dimensional roughness levels $10^{-3} – 10^{-2}$. Note that the simulations are for sphere–sphere interactions and provide only a crude estimate of the sphere–cylinder case.[29] On the other hand, separate profilometer measurements on square LEGO® posts shows that the pieces have roughness features on the order of 1 – 10 μm, indicating a comparable non-dimensional roughness for the cylinders. Finally, in Fig. 7, we also compare our macroscopic data with microfluidic *f*-DLD experiments obtained under similar conditions.[26] The microfluidic experiments were performed with silica microparticles that settled due to gravity in a sealed microfluidic device through a square array of cylindrical obstacles. We calculated $b_c$ from the reported critical angles at which the particles started to cross the columns within the micro-post array. By normalizing the data with respect to the radius of the cylindrical post, we can compare the microfluidic data with the macroscopic results reported here. There is one important difference with the experiments: the separation between posts is small, which could lead to a significant increase in the measured values for $b_c$. In addition, the roughness (or irregularities) of the microfabricated posts is significant compared to the macroscopic case. The agreement between the macroscopic and microscopic data is good given the differences discussed before.

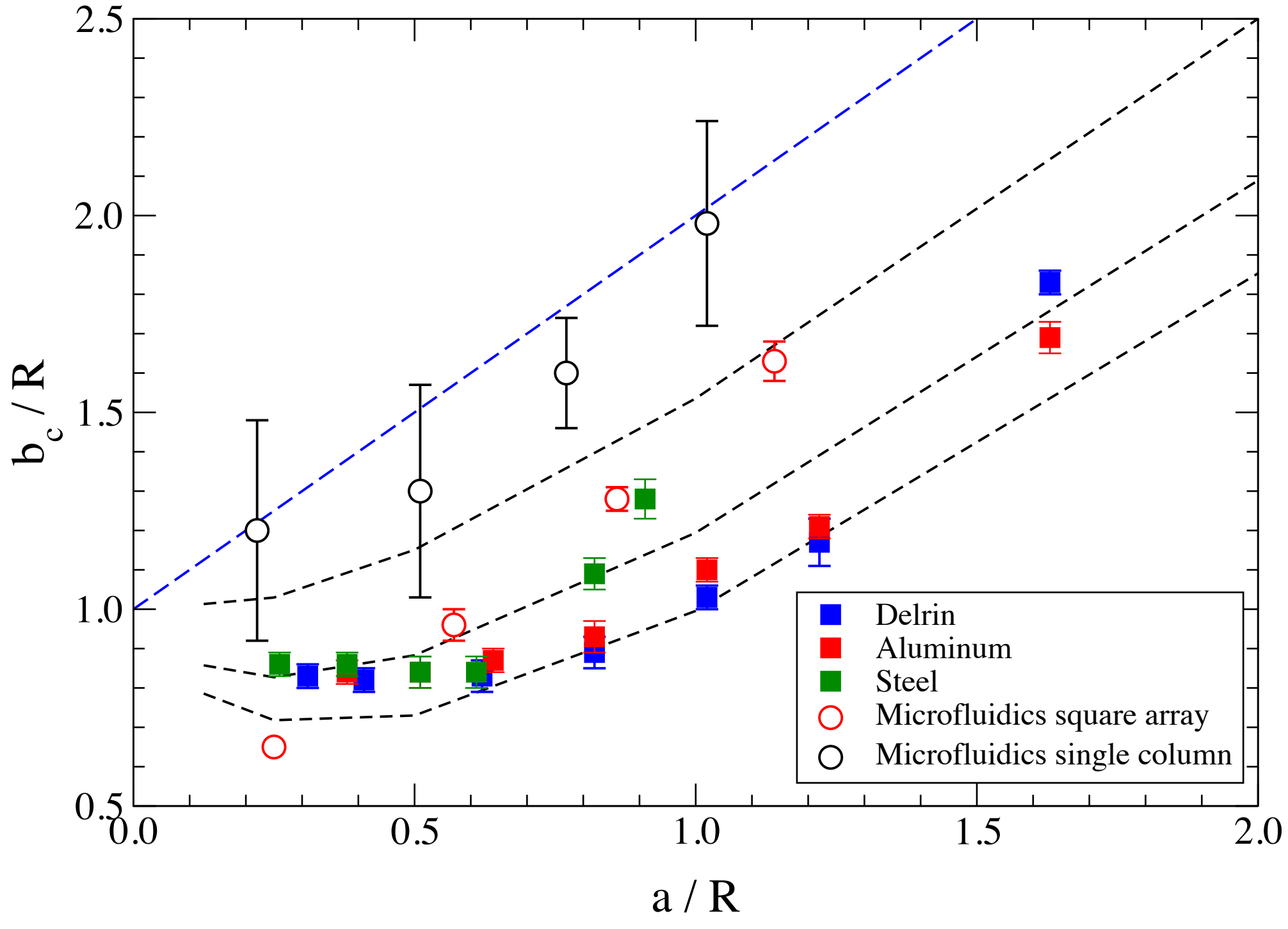

FIG. 7. Comparison of the data collected with microfluidic data in Ref [26] and Ref. [35] showing good agreement across different length scales. The dashed lines correspond to constant non-dimensional roughness $10^{-3}$, $10^{-2}$, and $10^{-1}$ from the bottom. The long-dashed line corresponds to $b_c = a + R$.

## IV. CONCLUSIONS

We showed a number of unique features in *f*-DLD using a macroscopic LEGO model. First, in the low Stokes numbers regime (St < 1), we have shown that *f*-DLD separation is a size-based phenomenon. In the intermediate Stokes number region (1 < St < 10) we demonstrate, for the first time, the capacity of a *f*-DLD system to separate particles by density. This is particularly important in light of the fact that in the low Stokes numbers regime, the critical offset parameter, and thus the migration angle, are remarkably independent of the material of the suspended particles. (Note that we investigated a number of different particles, including glass, PMMA, Nylon, Aluminum and Delrin ®.) We have also shown that the dimensionless critical offset as a function of the dimensionless particle size is a nearly universal curve. In fact, we have found good agreement between our data and results obtained in a microfluidic device that is three orders of magnitude smaller. As a result, the nearly universal curve found for the critical offset can be used, in principle, as a design curve that is independent not only of the scale of the system but also of the type of particles.

## V. APPENDIX: STUDIES OF GEOMETRY

In the dilute approximation for particle–obstacle collisions we assume that the particle reaches the asymptotic outgoing offset before it encounters the next obstacle, especially for the case of touching collisions and the determination of $b_c$. Working with Delrin® particles and 405 cSt silicone oil to minimize inertial effects, we present experiments in which the separation between obstacles in a given row has been varied.

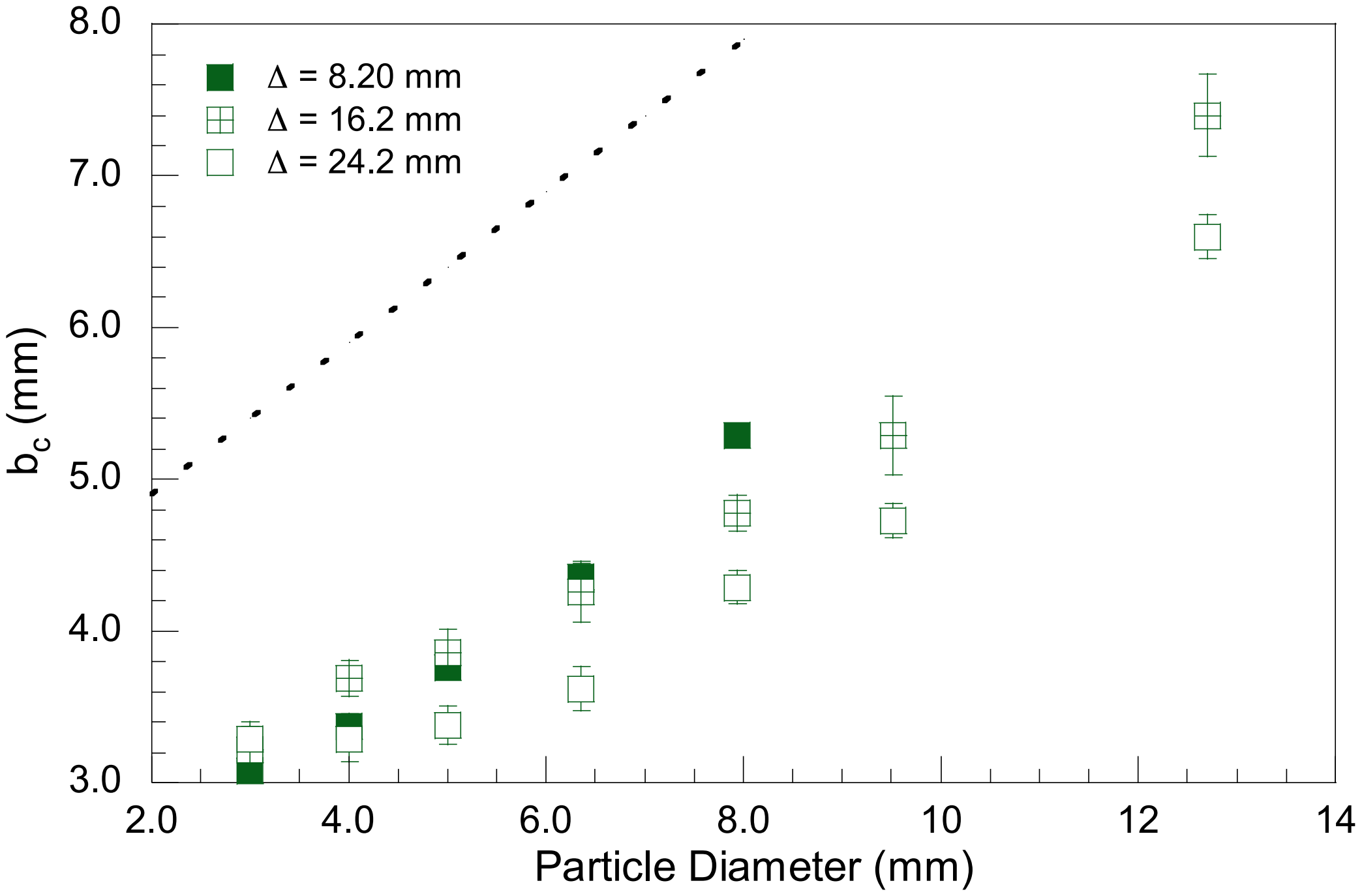

FIG. 8. Effect of the spacing between obstacles on the critical offset. The experiments were performed in silicone oil with Delrin particles. The dashed line corresponds to bc = a+R.

In Fig. 8 we present the critical offset as a function of particle size for obstacle-obstacle separations (Δ) of 8.2 mm, 16.2 and 24.2 mm (with obstacles R=3.9 mm). We observe that increasing the spacing between the obstacles reduces $b_c$. This is consistent with the fact that if a particle does not reach its asymptotic offset, the measured value for $b_c$ will be larger as the separation between obstacles is decreased. For *f*-DLD systems with closely spaced obstacles, the trajectory of the particle might not reach the critical offset due to additional hydrodynamic interactions with the neighbouring obstacle. Therefore the experimental $b_c$ value obtained from the measured critical angle is not the asymptotic value but instead corresponds to an effective critical offset that is dependent on the lattice spacing. Fortunately, this observation does not limit the utility of the model. Similar experiments featuring an array of closely spaced obstacles performed by Balvin *et al.*[1] show that an effective $b_c$ evaluated from the first critical angle can still predict the particle trajectory over the full range of forcing angles accurately.

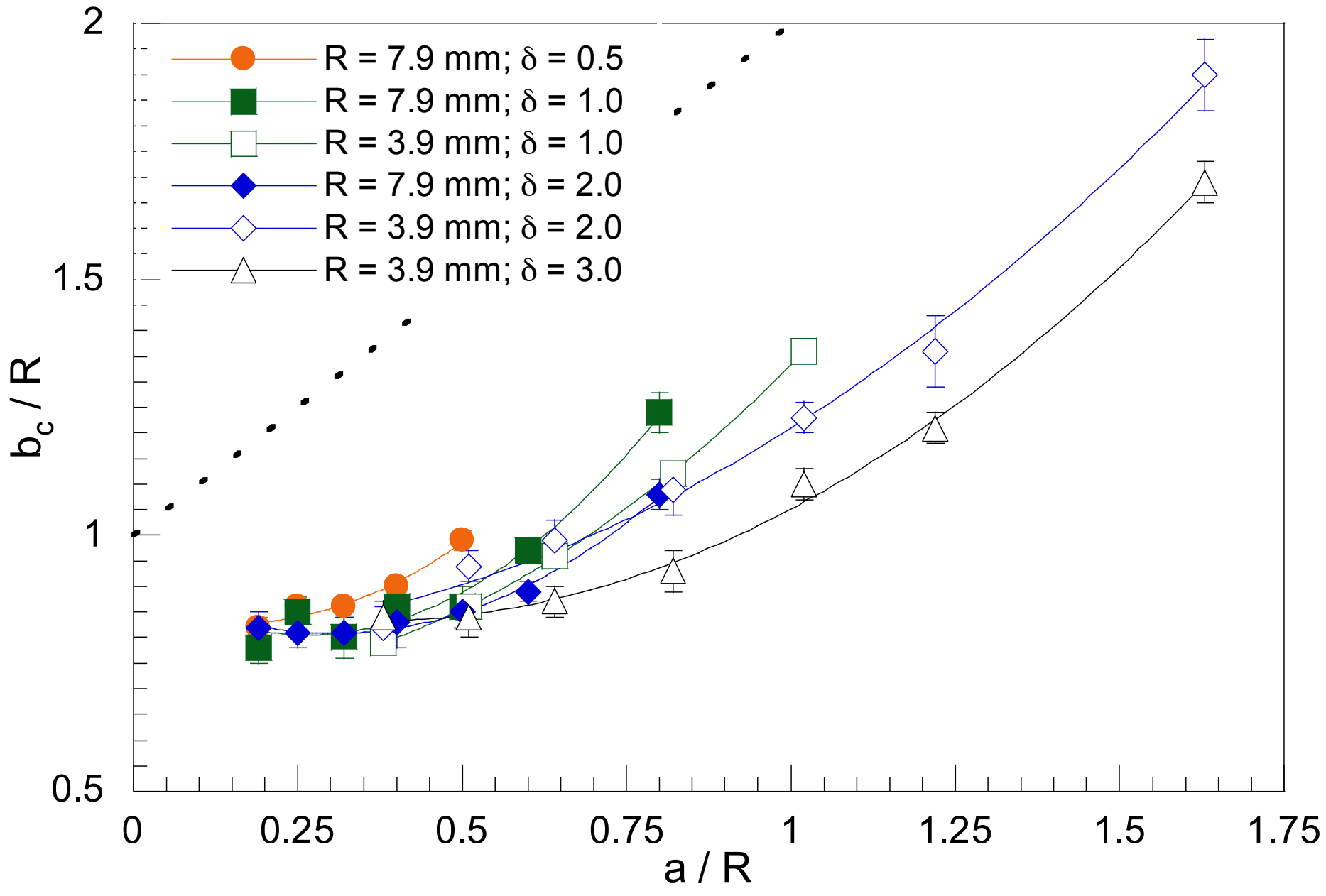

FIG. 9. Role of the aspect ratio on the critical offset for Delrin particles in silicone oil.

Finally, we performed experiments with larger obstacles while again varying the obstacle spacing so as to alter the system aspect ratio (δ), see Fig. 9. Consistent with our previous scaling results, we find that the data for different obstacle sizes is in good agreement when the aspect ratio δ is the same.

## VI. ACKNOWLEDGEMENTS

This material is based upon work partially supported by the National Science Foundation under Grant Nos. CBET-0933605, CMMI-0748094, and CBET-0954840.